\documentclass[twocolumn]{aastex62}

\newcommand{\kmps}{\,km\,s$^{\rm -1}$}

\newcommand{\Fe}{Fe {\sc i} 5435 \AA}
\newcommand{\Na}{Na {\sc i} }

\defcitealias{Cho2019}{I}

\received{April 1, 2018}
\shorttitle{Source depths of umbral oscillations}
\shortauthors{Cho et al.}
\begin{document}

\title{Source Depth of Three-minute Umbral Oscillations}

\author[0000-0001-7460-725X]{Kyuhyoun Cho}
\email{chokh@astro.snu.ac.kr}
\affil{Astronomy Program, Department of Physics and Astronomy, Seoul National University, Seoul 151-747, Republic of Korea}

\author[0000-0002-7073-868X]{Jongchul Chae}
\affil{Astronomy Program, Department of Physics and Astronomy, Seoul National University, Seoul 151-747, Republic of Korea}



\begin{abstract}
We infer the depth of the internal sources giving rise to three-minute umbral oscillations. Recent observations of ripple-like velocity patterns of umbral oscillations supported the notion that there exist internal sources exciting the umbral oscillations. We adopt the hypothesis that the fast magnetohydrodynamic (MHD) waves generated at a source below the photospheric layer propagate along different paths, reach the surface at different times, and excited slow MHD waves by mode conversion. These slow MHD waves are observed as the ripples that apparently propagate  horizontally. The propagation distance of the ripple given as a function of time is strongly related to the depth of the source. Using the spectral data of the \Fe\ line taken by the Fast Imaging Solar Spectrograph of the Goode Solar Telescope at Big Bear Solar Observatory, we identified five ripples and determined the propagation distance as a function of time in each ripple. From the model fitting to these data, we obtained the depth between 1000 km and 2000 km. Our result will serve as an observational constraint to understanding the detailed processes of magnetoconvection and wave generation in sunspots.
\end{abstract}


\section{Introduction} \label{sec:intro}
Oscillations of intensity and velocity are common in sunspots, in both umbrae and penumbrae.  At the chromospheric level, the oscillation periods are generally shorter than three-minutes in umbrae \citep{Bogdan2006}, and gradually increase with the distance from the sunspot center \citep{Nagashima2007}. Now it is generally accepted that sunspot oscillations are the slow magnetohydrodynamic (MHD) waves propagating along magnetic field lines \citep{Centeno2006, Jess2013}. Multi-line spectral observations indicated that the waves propagate upwardly inside umbrae \citep{Felipe2010}.

Although it is  known that the slow MHD waves propagate along the magnetic fields, sunspot oscillations are often observed to propagate across the magnetic fields. The most well-known phenomenon is running penumbral waves \citep{Zirin1972}. They are usually observed as horizontal propagating patterns with fast speed in penumbral regions. It is widely accepted that they are the apparent waves caused by the slow MHD waves propagating along the different inclined penumbral magnetic field lines \citep{Bloomfield2007, Lohner2015, Madsen2015}. The difference in the inclination among the penumbral magnetic field lines produces the difference in the path length, and, hence, the time lag at the detection layer that increases with the distance from the sunspot center. As a result, we observe successive arrivals of the slow MHD waves, which appears as the apparent horizontal propagation. This interpretation can explain both the horizontal propagations and period increase with distance from the sunspot center \citep{Jess2013}.

Interestingly, the horizontally propagating waves are found even in umbrae. \citet{RouppevanderVoort2003} found spreading bright arcs from Ca {\sc ii} H filtergram data, which represent the horizontally-propagating pronounced intensity oscillations known as umbral flashes. Velocity oscillations often form concentric ripple-like patterns crossing the umbrae transversely \citep{Zhao2015, Cho2019}. Even more complicated shapes such as spiral wave patterns also exhibit the horizontal propagation inside umbrae \citep{Sych2014, Su2016, Felipe2019, Kang2019}. These horizontally propagating patterns inside umbrae cannot be explained by the time lag due to the magnetic field inclination unlike the case of the running penumbral waves since their inclination angle is very close to zero or 180 degrees.

The horizontally propagating pattern of oscillations in umbrae can be explained if the source of wave excitation is located at a point inside the interior. In the high plasma $\beta$ environment of the interior, such a source excites fast MHD waves that propagate in all the directions \citep{Zhugzhda1982, Zhugzhda1984}. When they reach the photosphere, the time of arrival depends on the path and hence varies with the position of the photosphere. This introduces the time lag in the photosphere, which appears as the horizontally propagating pattern. Theoretical studies indicated that the fast MHD waves can generate the slow MHD waves in the photosphere via the fast-to-slow mode conversion \citep{Cally2001}. If this explanation is feasible the propagation aspect of the horizontally propagating patterns must be affected by the internal structures of sunspots where the fast MHD waves passed and especially the depth of the wave source.

In the present work, we infer the depth of the wave source from the observed horizontally propagating patterns of oscillations. This work is based on our recent study (\citealt{Cho2019}, hereafter Paper \citetalias{Cho2019}) on the excitation events of umbral oscillations. Among them, we choose five ripple-like patterns that show concentric shapes and propagate horizontally. We build an internal model of a sunspot to calculate the ray path of the fast MHD waves. Then the source depth is estimated from the model fitting of the propagating distance of a Doppler pattern determined as a function of time in each event. The obtained values of source depth are compared to previous reports, and we discuss the results in association with the origin of the three-minute umbral oscillations.

\section{Data and Analysis} \label{sec:anal}
We observed umbral oscillations around 21:00 UT on 2017 June 15 using the Fast Imaging Solar Spectrograph (FISS, \citealt{Chae2013}). The target was located in the leading sunspot of AR 12663 (25\arcsec, 205\arcsec) which was near the solar disk center. This observation data set is the same as the one used in Paper  \citetalias{Cho2019}.

The FISS provides four-dimensional (two spatial, spectral, and temporal dimensions) data in two bands simultaneously. In this observation, the pair of the \Fe\ band and the \Na D2 5890 band was chosen. We utilize only the \Fe\ line data that has the advantage in line-of-sight velocity measurements inside sunspots because of the zero Land\'e g factor (g = 0). The FISS observed 40\arcsec\ $\times$ 13\arcsec\ field of view at every 13 seconds. The details of the basic data processing were described by \citet{Chae2013}. We inferred the line of sight Doppler velocity at all the positions from the FISS \Fe\ spectra using the Gaussian core fitting. We also exploit the speckle-reconstructed TiO 7057 \AA\ broadband filter images \citep{Cao2010} to check the photospheric features of our region of interest. The TiO images were used as the reference for the alignment of the FISS data. For magnetic field information, we use Near-InfraRed Imaging Spectropolarimeter (NIRIS) data \citep{Cao2012}.

 We identified five ripple patterns from the three-minute filtered Doppler velocity movie. The identified ripple patterns constitute the event 1 and 2 in Paper \citetalias{Cho2019}. The event 1 and 2 has a peak of power at 3.2 minute, and 4.1 minute period, respectively. Both events had more than half of the oscillation power between 2 and 4 minute period. From the wavelet filtering data through the passband of periods between 2 and 4 minutes \citep{Torrence1998}, we found simple concentric shaped ripple patterns. 

We determined the oscillation center of each ripple and traced its propagation. The oscillation center was identified with the position of the peak Doppler signal at the very early phase as in Paper \citetalias{Cho2019}. By applying the azimuthal averaging, we obtained the Doppler velocity as a function of time $t$ and distance from the oscillation center $x$. The pattern of azimuthally averaged Doppler velocity $v_D(x, t)$ was modelled by the function 
\begin{equation}
v_{D}(x, t) = a_0(t) \sin{[a_1(t) x+a_2(t)]}+a_3(t).
\end{equation}
The coefficients $a_0(t), a_1(t), a_2(t)$, and $a_3(t)$ were determined from the model fit in the distance range of 0.5\arcsec\ to 4\arcsec\ at each time. The values of $a_1(t)$, and $a_2(t)$ were then used to determine the position of peak blueshift $x_b(t) = [3 \pi /2- a_2(t)]/a_1(t)$ at each time. For the redshift cases, we used a formula $x_r(t) = [\pi /2 - a_2(t)]/a_1(t)$.

\section{Model of Wave Propagation} \label{sec:model}

\begin{figure*}
\centering
\includegraphics[width=1 \textwidth,clip=]{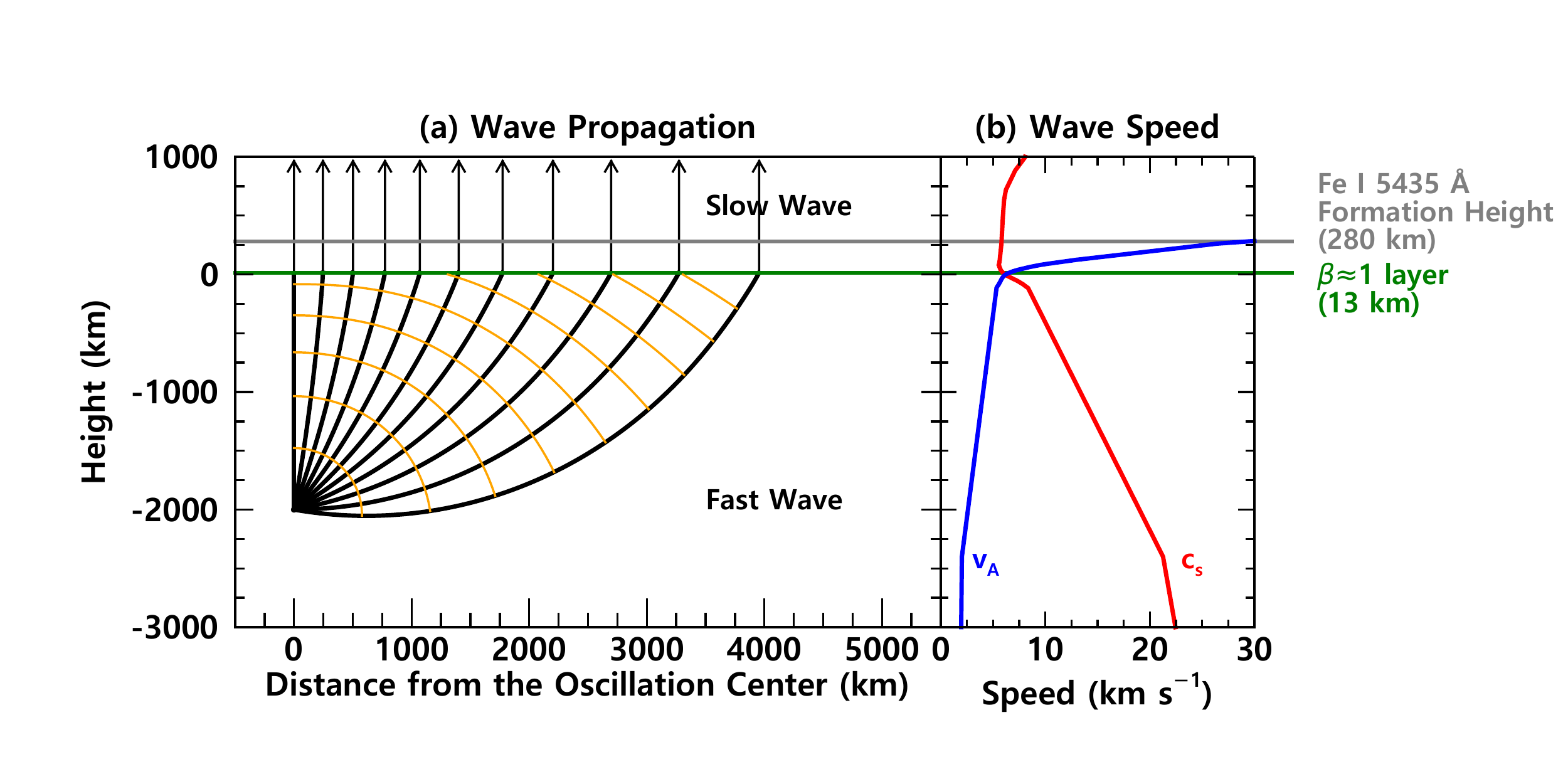}
\caption{(a) Wave propagation in solar interior and atmosphere when a source depth is 2000 km. Each black solid line indicates a ray path of the waves. Thick and thin black lines present the fast MHD waves and slow MHD waves, respectively. The orange solid lines indicate isochrone every 30 s. (b) Wave speed variation with height. Blue and green colors indicate Alfv\'en speed and sound speed, respectively. $\beta \simeq 1$ layer and the \Fe\ line formation height is indicated by a green and gray line, respectively. The animation shows two examples of ray paths of the fast MHD waves in the case of 500 km and 5000 km source depth. Red dots in the animation indicate the apparent waves at the plasma $\beta \simeq 1$ layer.} 
\label{fig1}
\end{figure*}

We adopted a model of wave propagation that takes into account several processes (See figure \ref{fig1}). We suppose a point-like event of wave generation takes place much below the umbral photosphere. The point source is in an environment of high plasma $\beta$, so the generated waves propagate mainly as the fast MHD waves. The propagation speed of the fast MHD waves $v_f$ is
\begin{eqnarray}\label{fast_speed}
v_f = \left ( {\frac{1}{2}  (c_s^2+v_A^2) + \sqrt{(c_s^2+v_A^2)^2-4 c_s^2 v_A^2 \cos^2 \theta} }\right )^{1/2}
 \end{eqnarray}
 where $c_s$ is sound speed, $v_A$ is Alfv\'en speed, $\theta$ is the angle between the propagating direction and the magnetic field lines. We assumed that magnetic fields in the umbra are vertical even below the photosphere. A part of the fast MHD waves propagate upward and reach the $\beta \simeq 1$ layer where the sound speed is roughly equal to the Alfv\'en speed, and where a part of fast MHD waves are converted to slow MHD waves \citep{Cally2001, Schunker2006}. These slow MHD waves propagate same distance along the vertical magnetic field lines. Thus, the time differences at the $\beta \simeq 1$ layer among different field lines are consequently preserved in the slow MHD wave mode. As a result, we come to observe the apparent horizontal propagation of waves at the \Fe\ line formation height of about 280 km \citep{Chae2017}.
 
We adopt the umbral E model of \citet{Maltby1986} to obtain the sound speed at each height. To determine the Alfv\'en speed, we need magnetic field strength as a function of height. We used the mean magnetic field strength of 2480 G inferred from the Milne-Eddington inverted Fe {\sc i} 1.56 $\mu$m NIRIS data. The formation height of that line in sunspot umbra is about 90 km \citep{Bruls1991}, and we employed the vertical gradient of -1 G/km \citep{Borrero2011}. By comparing the $c_s$ and $v_A$, we calculate the height where plasma $\beta$ is unity  (Figure \ref{fig1}b). The result is consistent with our picture that the deeper region shows high plasma $\beta$ and the upper region shows low plasma $\beta$. The $\beta \simeq 1$ layer is found to be about 13 km above the photospheric layer.

In reality, the fast MHD waves are refracted because of the variation of the phase velocity with heights. We calculate the refracted ray path of the fast MHD waves and the time of arrival on the $\beta \simeq 1$ layer using the eikonal method \citep{Weinberg1962, Moradi2008}. Given dispersion relation $D$, the relations between position $\mathbf{x}$, wave vector $\mathbf{k}$, time $t$, and frequency $\omega$ for the wave propagation of same phase are governed by following equations:

\begin{eqnarray}
\frac{d \mathbf{x}}{d \tau} & = & \frac{\partial D}{\partial \mathbf{k}} \\
\frac{d \mathbf{k}}{d \tau} & = & - \frac{\partial D}{\partial \mathbf{x}} \\
\frac{d t}{d \tau} & = & - \frac{\partial D}{\partial \omega} \\
\frac{d \omega }{d \tau} & = & - \frac{\partial D}{\partial \tau}
\end{eqnarray}

In case of the fast MHD waves, the dispersion relation is given by
\begin{eqnarray}
D = \omega^4 - (c_s^2 + v_A^2) k^2 \omega^2 + c_s^2 v_A^2 k^2 k_z^2 = 0 .
\end{eqnarray}
We set the oscillation frequency $\omega$ to be 0.035 $s^{-1}$, which corresponds to about 3 minutes. In fact, our results do not depend on the frequency because we do not take into account gravity \citep{Kalkofen1994, Chae2015}. As it is not a dispersive medium, phase speed of the fast MHD waves does not depend on the frequency (see Equation \ref{fast_speed}). The ray paths of the fast MHD waves are calculated from a point with a given depth for several initial angles (Figure \ref{fig1}a). As a result, we obtain the distance from the oscillation center and duration until the fast MHD waves reach the $\beta \simeq 1$ layer and then we are able to calculate the horizontal position of the apparent wave as a function of time. 

Interestingly, the speed of apparent horizontal propagation strongly depends on the depth of the source. The animation associated with Figure \ref{fig1} clearly shows that a shallower source yields a slower apparent speed while a deeper source yields a faster apparent speed. This is because ray paths of the waves originated from a deeper source are not much different from each other. So, the fast MHD waves arrive almost at the same time, and it looks like that the apparent waves propagate quickly. We can explain the variation of the propagating speed with the distance from the oscillation center in a similar way. We estimate the depth by fitting the observed distance of the peak blueshift $x_b(t)$ by the one calculated with the model of wave propagation. 

\section{Results} \label{sec:result}

\begin{figure*}
\centering
\includegraphics[width= \textwidth,clip=]{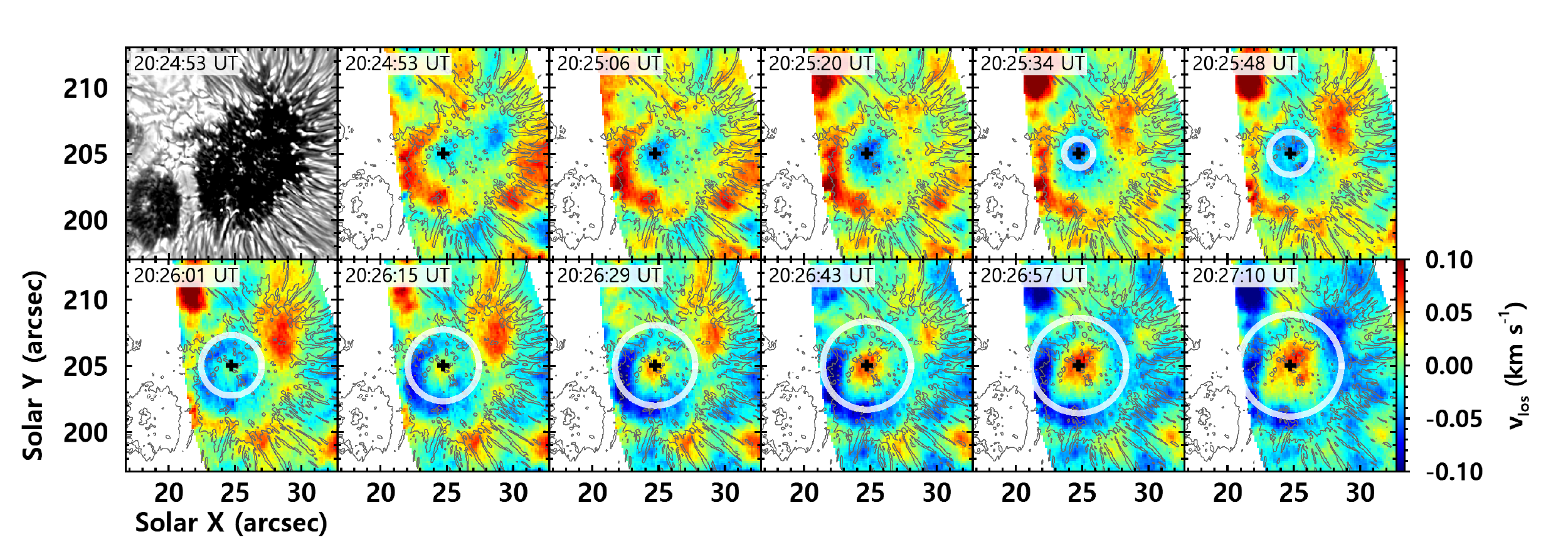}
\caption{An example of TiO broadband filter image and time series of the FISS three-minute filtered Doppler maps for ripple 1. The black cross symbol represents the position of the oscillation center, and the white circles indicate the determined positions of ripple 1. The radii of the white circles are distances of the propagating ripple 1 from the oscillation center, which are determined by sinusoidal function fitting (See Figure \ref{fig3}). Black contours represent the umbral-penumbral boundary and the positions of umbral dots. The animation shows all five ripples in the same way.}
\label{fig2}
\end{figure*}

Figure \ref{fig2} shows the time series of the Doppler velocity maps showing the propagation of ripple 1. We found that a blueshift ripple emerged from the oscillation center near the umbral center; then it propagated radially. The associated animation clearly shows that the selected five ripples propagated outward with concentric circle shape.

\begin{figure*}
\centering
\includegraphics[width=0.8 \textwidth,clip=]{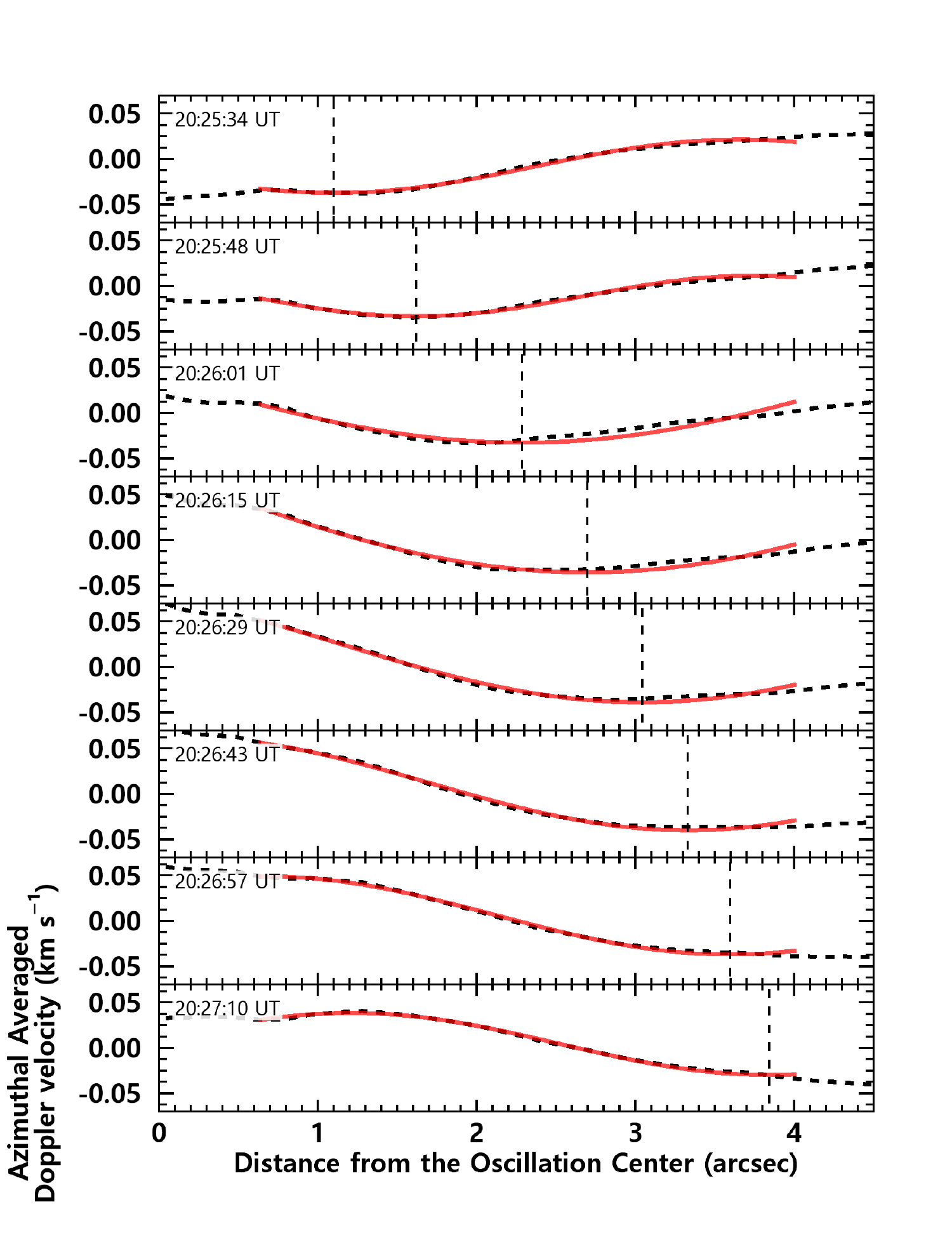}
\caption{An example of the determination of the position of ripple 1. Rows exhibit different observation times. the black dashed line is azimuthal averaged Doppler velocity with distance from the oscillation center. The red solid line indicates the sinusoidal fitting results. The vertical dashed line presents the determined position of ripple 1.}
\label{fig3}
\end{figure*}

We measured the distances of ripple 1 from the oscillation center (Figure \ref{fig3}). The pattern of azimuthally-averaged Doppler velocity oscillated with an amplitude of about 0.05 \kmps\ and propagated outward. The fitting results obviously show that the pattern can be well described by a sinusoidal function of distance. We will describe ripple 1 in detail. From the fitting, we find that the amplitude of Doppler velocity range from 0.023 to 0.11 \kmps\ with a mean of about 0.048 \kmps. The spatial wavelength was determined to range from 4.2\arcsec\ (3000 km) to 11\arcsec\ (7800 km) with the mean value of 6.8\arcsec\ (4900 km). The ripple moved away from the oscillation center as we expected. Moreover, the distance between two successive ripples decreased with time, which implies that the speed decreased with the propagation distance. The determined speed was initially 27 \kmps\ and decreased to about 13 \kmps, which is much faster than the slow MHD waves at the detection layer.

\begin{figure*}
\centering
\includegraphics[width=1 \textwidth,clip=]{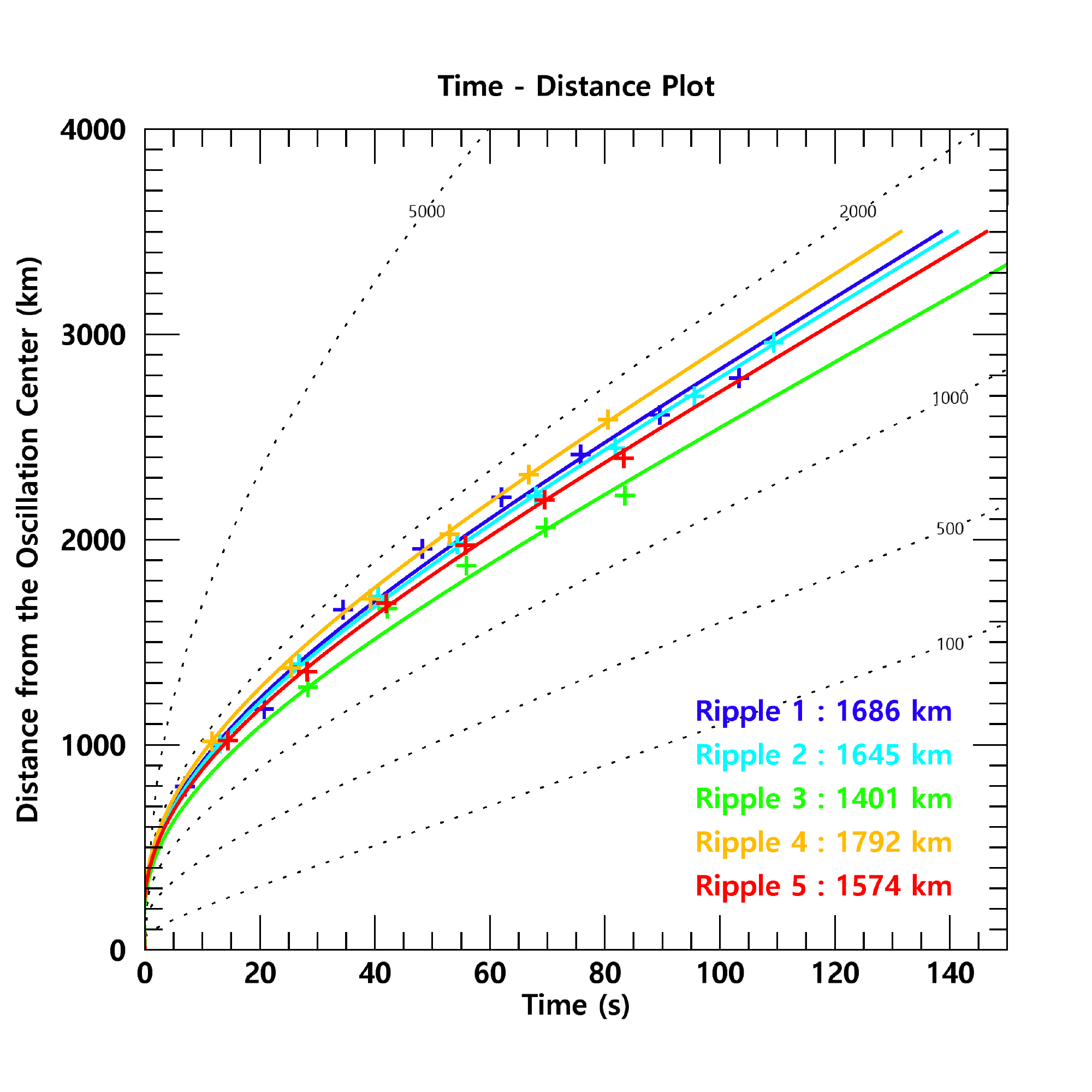}
\caption{Time-Distance plot for the five ripples. The cross symbols and solid line present the determined positions of the ripples from the observations and the fitting result from apparent wave calculations, respectively. Colors indicate different ripples. The estimated source depths are presented in the lower right corner. Dashed lines represent the result of apparent waves from the model calculations with 100, 500, 1000, 2000, 5000 km source depth. }
\label{fig4}
\end{figure*}

The dashed lines in Figure \ref{fig4} shows the distance of the ripple as a function of time calculated using our model of wave propagation in each case of the source depths. In each time-distance plot, the horizontal propagation speed can be inferred from the slope of the curve. It usually starts with a high value and decreases with time, and approaches an asymptotic value. This asymptotic speed as well as the average speed depends on the model and increases with the source depth of the model.

The data (distances and times) of all the ripple marked in the figure by the symbols closely match the models presented in thick solid curves. The data are very similar to the models particularly in the trend of the decreasing velocity with the time or the propagation distance from the oscillation center. Based on our results, we conclude that the depth of the oscillation sources ranges from about 1400 km to 1800 km with a mean value of about 1600 km below the photosphere.

\section{Discussion} \label{sec:diss}

We observed the horizontally propagating ripples in a sunspot umbra. These cannot be explained by slow MHD waves only, because slow MHD waves propagate vertically along magnetic fields. We attempted to interpret their horizontal propagations as the apparent ones caused by the time lag. It is assumed that the time lag is the result of the different arrival times of the fast MHD waves below the plasma $\beta \simeq 1$ layer. We constructed a model based on this scenario and fitted the observational data. The observational results were successfully reproduced by our calculations with a depth of about 1600 km. The decrease of the propagating speed with distance is also adequately explained by our model calculations.

We expect that there are two major sources of error in calculating the source depth. First, the measurement error of the positions of ripples can affect the result. This is also closely related to the determination of the oscillation center. We conjecture that the measurement error is less than 1\arcsec\ which corresponds to about 1000 km variation of depth. The second source is an inaccurate model. We deduced the variation of the sound speed and Alfv\'en speed with depth from the sunspot atmospheric model and the magnetic field information. The most ambiguous information is the vertical gradient of the magnetic field strength. In the high $\beta$ regime, however, the fast MHD waves are not significantly affected by the magnetic field strength. We have tested several cases of magnetic field strength, then concluded that there might be an error of about a few hundred km in the depth estimation. Thus it is reasonable to state that the wave source is roughly located between 1000 km and 2000 km depth. To obtain more accurate depth, and explain the frequency dependent behavior of the waves \citep{Zhao2016}, it is necessary to examine the effects of gravity on the wave propagation. 

\begin{figure*}
\centering
\includegraphics[width=1 \textwidth,clip=]{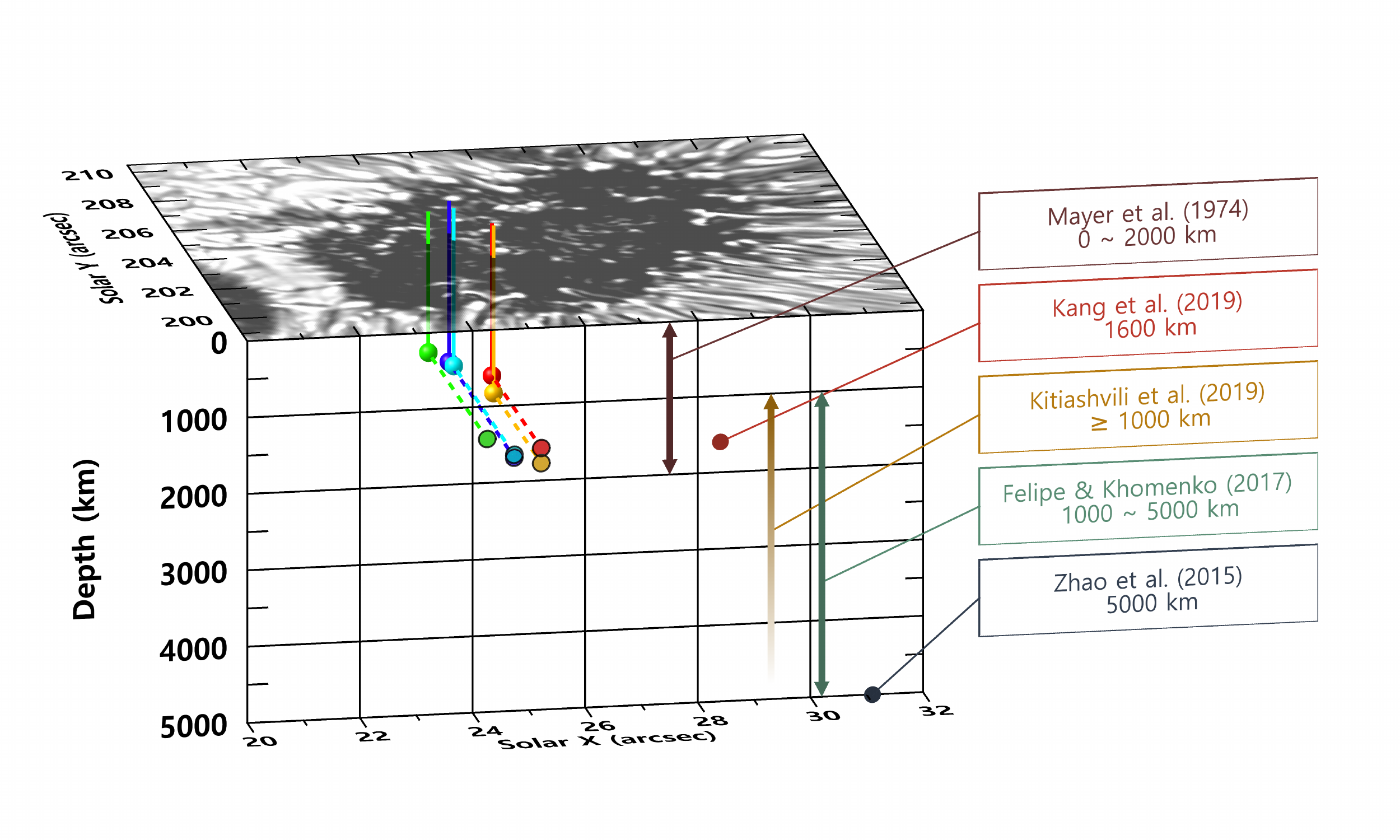}
\caption{Source positions of the three-minute umbral oscillations in 3-D map and comparison with previous studies. Small spheres present the source position. The same colores with Figure \ref{fig4} used to indicate the results from the different ripples. Vertical solid lines are auxiliary lines that indicate the positions of the oscillation centers in the plane of the sky. Corresponding colored circles are the projected positions of the oscillation centers on the xz plane. Previous studies are presented in the right. }
\label{fig5}
\end{figure*}

Our result is in agreement with previous studies (Figure \ref{fig5}). \citet{Meyer1974} theoretically studied the instability of the sunspot model. As a result, it was demonstrated that overstable oscillation may occur in the top 2000 km of sunspots with the parallel motion to the magnetic field. From the HMI Dopplergram data, \citet{Zhao2015} found fast-moving waves with speeds of about 45 \kmps\ using the time-distance cross-correlation method. They conjectured that a disturbance occurring at about 5000 km under the sunspot surface from the MHD sunspot model and the ray-path approximation with magnetic fields. \citet{Feilpe2017} performed MHD numerical simulations to confirm the dependence of the source depth. They concluded that the measured horizontal fast-moving waves consistent with waves generated between about 1000 km and 5000 km beneath the sunspot photosphere. Analyzing the result of the 3-D radiative MHD simulations, \citet{Kitiashvili2019} identified that most of the wave sources are concentrated below 1000 km in the pore-like magnetic structure. \citet{Kang2019} argued that 1600 km of source depth is required to explain the spiral wave patterns in sunspots using a simplified wave propagation model. Our estimate, the source depth of about 1600 km, is fairly consistent with these results.

Our study contributes to the understanding of the generation of the umbral three-minute oscillations. Our result suggests that the origin of the umbral three-minute oscillations is located below the photosphere as a point source. It supports the internal excitation as the origin of the umbral oscillations. Paper \citetalias{Cho2019} found the association of the oscillation center and the umbra dots in the photosphere. Considering that umbra dots are regarded as the signature of the magnetoconvection, we can conjecture the shape of a magnetoconvection cell. The average size of umbral dots is less than 1\arcsec ($\simeq$ 700 km). Taking into account the source depth close to 2\arcsec\ ($\simeq$ 1500 km), we imagine that the cell of the magnetoconvection may be vertically elongated. This vertical elongation seems to be reasonable in the umbral environment permeated by strong vertical magnetic field lines. Previous studies based on MHD simulations exhibited such vertically elongated convection cells with comparable physical size \citep{Schussler2006, Rempel2009}.

It would be worthwhile to examine the azimuthal dependence of the horizontal propagation. We only used the azimuthally averaged Doppler velocity in this study. It is known that the asymmetric behavior arises from the difference of the time legs, which is affected by a different path, magnetic field inclination, flows, and temperature variation (e.g. \citealt{Kosovichev1997}). Therefore, the relevant studies will provide information about the magnetoconvection inside sunspot, and eventually the internal structure of sunspots.

%It suggests the possibility that the discovered event is a group of individual events.   individual structure : next generation higher resolution.

% similar timescale and length scale. -> It may exist a magnetoconvection cell which is resemble to the granule in the Quiet sun.

% 10^6 enregy flux. similare results from previouse studies. enough to heat the higher atmosphere. (100 W/m)

% one acoustic event can generate 10^25 erg energy. comparable amount of the energy with nanoflares (Krucker and Benz, 1998)(Benz and Krucker (2002

\acknowledgments
This work was supported by the National Research Foundation of Korea (NRF-2017R1A2B4004466).


\begin{thebibliography}{}

\bibitem[Bruls et al.(1991)]{Bruls1991} Bruls, J. H. M. J., Lites, B. W., \& Murphy, G. A. 1991, Solar Polarimetry,
ed. L. J. November (Sunspot : National Solar Observatory), 444

\bibitem[Bloomfield et al.(2007)]{Bloomfield2007} Bloomfield, D.~S., Lagg, A., \& Solanki, S.~K.\ 2007, \apj, 671, 1005

\bibitem[Bogdan \& Judge(2006)]{Bogdan2006} Bogdan, T.~J., \& Judge, P.~G.\ 2006, Philosophical Transactions of the Royal Society of London Series A, 364, 313

\bibitem[Borrero, \& Ichimoto(2011)]{Borrero2011} Borrero, J.~M., \& Ichimoto, K.\ 2011, Living Reviews in Solar Physics, 8, 4

\bibitem[Cally(2001)]{Cally2001} Cally, P.~S.\ 2001, \apj, 548, 473

\bibitem[Cao et al.(2010)]{Cao2010} Cao, W., Gorceix, N., Coulter, R., et al.\ 2010, \procspie, 7735, 77355V

\bibitem[Cao et al.(2012)]{Cao2012} Cao, W., Goode, P.~R., Ahn, K., et al.\ 2012, Second ATST-EAST Meeting: Magnetic Fields from the Photosphere to the Corona., 463, 291

\bibitem[Centeno et al.(2006)]{Centeno2006} Centeno, R., Collados, M., \& Trujillo Bueno, J.\ 2006, \apj, 640, 1153

\bibitem[Chae et al.(2013)]{Chae2013} Chae, J., Park, H.-M., Ahn, K., et al.\ 2013, \solphys, 288, 1

\bibitem[Chae \& Goode(2015)]{Chae2015} Chae, J., \& Goode, P.~R.\ 2015, \apj, 808, 118

\bibitem[Chae et al.(2017)]{Chae2017} Chae, J., Lee, J., Cho, K., et al.\ 2017, \apj, 836, 18

\bibitem[Cho et al.(2019)]{Cho2019} Cho, K., Chae, J., Lim, E.-. kyung ., et al.\ 2019, \apj, 879, 67

\bibitem[Felipe et al.(2010)]{Felipe2010} Felipe, T., Khomenko, E., Collados, M., \& Beck, C.\ 2010, \apj, 722, 131

\bibitem[Felipe \& Khomenko(2017)]{Feilpe2017} Felipe, T., \& Khomenko, E.\ 2017, \aap, 599, L2

\bibitem[Felipe et al.(2019)]{Felipe2019} Felipe, T., Kuckein, C., Khomenko, E., et al.\ 2019, \aap, 621, A43

\bibitem[Jess et al.(2013)]{Jess2013} Jess, D.~B., Reznikova, V.~E., Van Doorsselaere, T., Keys, P.~H., \& Mackay, D.~H.\ 2013, \apj, 779, 168

\bibitem[Kalkofen et al.(1994)]{Kalkofen1994} Kalkofen, W., Rossi, P., Bodo, G., et al.\ 1994, \aap, 284, 976

\bibitem[Kang et al.(2019)]{Kang2019} Kang, J., Chae, J., Nakariakov, V.~M., et al.\ 2019, \apjl, 877, L9

\bibitem[Khomenko, \& Collados(2015)]{Khomenko2015} Khomenko, E., \& Collados, M.\ 2015, Living Reviews in Solar Physics, 12, 6

\bibitem[Kitiashvili et al.(2019)]{Kitiashvili2019} Kitiashvili, I.~N., Kosovichev, A.~G., Mansour, N.~N., et al.\ 2019, \apj, 872, 34

\bibitem[Kosovichev, \& Duvall(1997)]{Kosovichev1997} Kosovichev, A.~G., \& Duvall, T.~L.\ 1997, Score'96 : Solar Convection and Oscillations and Their Relationship, 241

\bibitem[L{\"o}hner-B{\"o}ttcher \& Bello Gonz{\'a}lez(2015)]{Lohner2015} L{\"o}hner-B{\"o}ttcher, J., \& Bello Gonz{\'a}lez, N.\ 2015, \aap, 580, A53

\bibitem[Madsen et al.(2015)]{Madsen2015} Madsen, C.~A., Tian, H., \& DeLuca, E.~E.\ 2015, \apj, 800, 129

\bibitem[Maltby et al.(1986)]{Maltby1986} Maltby, P., Avrett, E.~H., Carlsson, M., et al.\ 1986, \apj, 306, 284

\bibitem[Meyer et al.(1974)]{Meyer1974} Meyer, F., Schmidt, H.~U., Weiss, N.~O., \& Wilson, P.~R.\ 1974, \mnras, 169, 35

\bibitem[Moradi \& Cally(2008)]{Moradi2008} Moradi, H., \& Cally, P.~S.\ 2008, \solphys, 251, 309

\bibitem[Nagashima et al.(2007)]{Nagashima2007} Nagashima, K., Sekii, T., Kosovichev, A.~G., et al.\ 2007, \pasj, 59, S631

\bibitem[Rempel et al.(2009)]{Rempel2009} Rempel, M., Sch{\"u}ssler, M., \& Kn{\"o}lker, M.\ 2009, \apj, 691, 640

\bibitem[Rouppe van der Voort et al.(2003)]{RouppevanderVoort2003} Rouppe van der Voort, L.~H.~M., Rutten, R.~J., S{\"u}tterlin, P., Sloover, P.~J., \& Krijger, J.~M.\ 2003, \aap, 403, 277

\bibitem[Schunker, \& Cally(2006)]{Schunker2006} Schunker, H., \& Cally, P.~S.\ 2006, \mnras, 372, 551

\bibitem[Sch{\"u}ssler \& V{\"o}gler(2006)]{Schussler2006} Sch{\"u}ssler, M., \& V{\"o}gler, A.\ 2006, \apjl, 641, L73

\bibitem[Su et al.(2016)]{Su2016} Su, J.~T., Ji, K.~F., Cao, W., et al.\ 2016, \apj, 817, 117

\bibitem[Sych, \& Nakariakov(2014)]{Sych2014} Sych, R., \& Nakariakov, V.~M.\ 2014, \aap, 569, A72


\bibitem[Torrence \& Compo(1998)]{Torrence1998} Torrence, C., \& Compo, G.~P.\ 1998, Bulletin of the American Meteorological Society, 79, 61

\bibitem[Weinberg (1962)]{Weinberg1962} Weinberg, Steven\ 1962, \pra, 126, 1899

\bibitem[Zhao et al.(2015)]{Zhao2015} Zhao, J., Chen, R., Hartlep, T., \& Kosovichev, A.~G.\ 2015, \apjl, 809, L15

\bibitem[Zhao et al.(2016)]{Zhao2016} Zhao, J., Felipe, T., Chen, R., et al.\ 2016, \apjl, 830, L17

\bibitem[Zhugzhda \& Dzhalilov(1982)]{Zhugzhda1982} Zhugzhda, I.~D., \& Dzhalilov, N.~S.\ 1982, \aap, 112, 16

\bibitem[Zhugzhda \& Dzhalilov(1984)]{Zhugzhda1984} Zhugzhda, I.~D., \& Dzhalilov, N.~S.\ 1984, \aap, 133, 333

\bibitem[Zirin \& Stein(1972)]{Zirin1972} Zirin, H., \& Stein, A.\ 1972, \apjl, 178, L85
\end{thebibliography}
\end{document}